\documentclass[10pt,a4paper]{revtex4}
\usepackage[dvips]{graphicx}

\begin{document}
\title{A microscopic approach to critical phenomena at interfaces: an application to complete wetting in the Ising model}
\author{A. Orlandi$^{1,4,}$}\email{orlandi@pv.infn.it}\author{A. Parola$^{2,4}$}\author{L. Reatto$^{3,4}$}
\address{$^1$ {\em Dipartimento di fisica Nucleare e Teorica, Universit\`a di Pavia, Via Bassi 6, 27100 Pavia, Italy}\\
$^2$ {\em Dipartimento di Fisica e Matematica, Universit\`a dell'Insubria, 
Via Valleggio 11, 22100  Como, Italy }\\
$^3$ {\em Dipartimento di Fisica,  Universit\`a di  Milano, Via Celoria 16,Milano, Italy} \\              
$^4$ {\em Istituto Nazionale per la  Fisica della Materia}                                  
}                
\begin{abstract}
We study  how the formalism of the  Hierarchical Reference Theory (HRT) can be extended to 
inhomogeneous systems. HRT is a liquid state theory which implements the basic ideas of 
Wilson momentum shell renormalization group (RG) to microscopic Hamiltonians. In the case of 
homogeneous systems, HRT provides accurate results even in the critical region, where it reproduces
scaling and non-classical critical exponents. We applied the HRT to study wetting critical 
phenomena in a planar geometry. Our formalism avoids the explicit definition 
of effective surface Hamiltonians but leads, close to the wetting transition, 
to the same renormalization group equation already studied by RG techiques. However,
HRT also provides information on the non universal quantities because it does not require
any preliminary coarse graining procedure. A simple approximation to the infinite HRT
set of equations is discussed.
The HRT evolution equation for the surface free energy is numerically integrated in a 
semi-infinite three-dimensional Ising model and the complete wetting phase transition
is analyzed. A renormalization of the adsorption critical amplitude and of the wetting parameter
is observed. Our results are compared to available Monte Carlo simulations.
\end{abstract}
\maketitle

\section{Introduction}
Effective interfacial Hamiltonians are widely used to describe the large-scale fluctuations 
that occur near surface critical phenomena such as wetting \cite{wetting}. 
Of course these models are not truly microscopic but are usually considered valid 
for length scales larger than some appropriate cut-off. The prevailing belief is that 
interfacial models may be derived from more microscopic approaches if the bulk degrees 
of freedom are integrated out. Needless to say this is an extremely difficult task 
and interfacial models still retain a partly phenomenological status. Therefore it 
is interesting to develop a genuine microscopic approach to study interfacial behaviors. 
Effective Hamiltonians are used to describe the critical behavior also in homogeneous systems
because they allow a direct implementation of renormalization group (RG) ideas.  
For such bulk systems, the Hierarchical Reference Theory \cite{HRT,HRT1} provides a systematic 
way to derive an effective Hamiltonian from a given microscopic model \cite{Brognara}. 
This theory, which implements the basic ideas of Wilson momentum space renormalization group 
for microscopic Hamiltonians, allows to derive an exact hierarchy of differential equations 
describing the evolution of the free energy and of the $n$-point correlation functions of the 
system when fluctuations on larger and larger length-scales are included. This hierarchy 
of differential equations can be closed, for example, by imposing an approximation for the pair correlation 
function, usually of the Ornstein-Zernike (OZ) form.
Already at this level of approximation, the HRT shows genuine non classical critical behavior. 

The extension of the HRT formalism to inhomogeneous systems would enable us to study the 
interfacial phase transition starting from the microscopic model, without the 
explicit introduction of effective interfacial Hamiltonians. Given two homogeneous phases, 
$\alpha$ and $\beta$, in contact with a third passive phase (a wall), we study the particular 
case of complete wetting in planar geometry \cite{wetting}, {\em i.e.} the phase transition
which corresponds to the growth of an infinitely thick liquid layer close to the wall
when bulk phase coexistence is approached at fixed temperature. The standard 
fluctuation theory  of the wetting transition, beyond the mean field approximation, 
is described by using  the capillary wave (CW) effective Hamiltonian in $d$ dimensions\cite{Forgacs,Boulter}
\begin{equation}
  H_{CW}=\int d^{d-1}\mathbf{x}\;\left[\frac{\Sigma}{2}\big(\nabla l(\mathbf{x})\big)^2+W(l(\mathbf{x}))\right] \label{CW}
\end{equation}
where $l(\mathbf{x})$ is a collective coordinate which represents the distance of the 
$\alpha-\beta$ interface from the wall, $W(l)$ is the effective interface potential 
which describes the effective interaction between the wall and the $\alpha-\beta$ interface 
and $\Sigma$ is the stiffness coefficient of the interface. For isotropic models $\Sigma$ 
can be identified with the surface tension, but 
it may also depend on the curvature of the interface. The form of effective interfacial 
potential $W(l)$ depends on the range of the  microscopic interaction. In the case of short 
range potentials, for large $l$ it is parametrized as
\cite{Forgacs,Boulter,fisher,lipowsky,lipowsky1,Jin}:
\begin{equation}
W(l)\simeq -A\exp(-ml/\xi)+B\exp(-nl/\xi) \label{efpotential}
\end{equation}
where $m$, $n$ are two dimensionless constants with $n>m$. The parameter $\xi$ can be 
identified with the bulk correlation length. It is important to note that the use of 
such an effective surface Hamiltonian can be justified only when bulk fluctuations are 
small {\em i.e.} far for bulk critical point \cite{Huse}. In three dimension,
renormalization group studies on the Hamiltonian (\ref{CW}) \cite{lipowsky,fisher} 
predict that the wetting critical behavior is non-universal depending on the value of the 
wetting parameter:
\begin{equation}
\omega=\frac{k_{B}T}{4\pi\Sigma\xi_b^2} \label{w}
\end{equation}
where $\xi_b$ is the {\em true} bulk correlation length which governs the exponential decay 
of correlations in real space \cite{EHP,Tarko}. At the critical wetting transition 
({\em i.e.} wetting at coexistence) the critical exponents depend on 
this parameter \cite{fisher} while at complete wetting ({\em i.e.} wetting approaching coexistence) 
 the critical exponents are predicted to remain 
mean field like but critical amplitudes are $\omega$ dependent. For example 
the interface height $l$ grows as: 
\begin{equation}
\frac{l}{\xi_b}\sim \left(1+\frac{\omega}{2}\right)\log\big(\Delta \mu\big) \label{amplitude}
\end{equation}
where $\Delta\mu=\mu-\mu_0$ is the deviation of the chemical potential from the value at coexistence. The mean field 
approximation to an effective $\phi^4$ hamiltonian provides a value $\omega_{mf}=0$ via Eq.(4).  

The three-dimensional semi-infinite Ising model is one of the simplest microscopic models 
which may be simulated to test the RG predictions. The value of the wetting parameter $\omega$ 
is a function of the temperature and has been theoretically estimated \cite{Wen,EHP}: 
$\omega\sim 0.8$ for $T_c>T\geq 0.6T_c$. 
Extensive Monte Carlo \cite{Binder} simulation studies appear to be consistent with 
$\omega\sim 0.3$ at $T_{sim}\simeq 0.663 T_c$ \cite{BinderParry} a much smaller value than the predicted one. 
One possible solution of this discrepancy between the simulations and the RG results is the 
inadequacy of the CW Hamiltonian and the introduction of  more general effective interfacial Hamiltonians 
\cite{Parry1,Parry2,Parry3,Parry4}.

The HRT theory, suitably generalized to deal with inhomogeneous systems, 
can be used to study the effects of thermal fluctuations 
beyond  the mean field behavior without reference to the CW approach 
and so without introducing $\omega$ as an external parameter. 
We first obtain the HRT surface evolution equation for the Ising model which is then
numerically solved in the case of complete wetting in three dimensions. The divergence of 
the adsorption $\Gamma$ is studied when coexistence is approached and the 
critical amplitude is evaluated and compared to simulations.
This work is organized as follows: in Section II we extend HRT approach to the case 
of inhomogeneous systems with planar geometry. We derive the evolution equation for 
the bulk and surface free energy. We also show that in the asymptotic region the HRT 
equations reduce to the known RG approach. In Section III we study complete wetting 
for a lattice gas model with nearest-neighbor interactions (which is equivalent to the Ising model). 
We investigate how fluctuations modify the mean field picture by integrating the HRT 
surface equation. In Section IV we briefly summarize the most relevant results and compare them
to the avaliable simulation data. 

\section{HRT for Inhomogeneous systems}
\subsection{Evolution equations}
The starting point in the derivation of HRT equations is the separation of 
the inter-atomic potential $v(r)$ in two parts:
\begin{equation}
v(\mathbf{r})=v_R(\mathbf{r})+w(\mathbf{r})
\end{equation}
where $v_R(\mathbf{r})$ is the short-range repulsive part of $v(\mathbf{r})$. 
The thermodynamic and structural properties of the system with interaction 
$v_R(\mathbf{r})$, the ``reference system'', are considered known, at least numerically. 
It is also assumed that there is no phase transition in the reference system. 
Instead $w(\mathbf{r})$ is a (mostly) attractive term, which triggers the liquid-vapor phase 
transition. Using this separation and performing a  Legendre transformation on the grand 
partition function, a formal diagrammatic expansion for the  Helmholtz free energy 
can be written to all orders in perturbation theory \cite{HRT1}. We implement the basic ideas 
of Wilson's RG approach \cite{Wilson} within such a formal perturbative expansion 
to study how the bulk and surface thermodynamic quantities evolve due to the inclusion of 
fluctuations.  This can be done by introducing a sequence of intermediate potentials 
characterized by an infrared cut-off in Fourier space, depending on a parameter $Q$. In the 
HRT for (off-lattice) bulk systems this cut-off is spherical in Fourier space, thereby 
respecting the isotropy of the interaction. Here we want to consider the case of 
inhomogeneous systems in the presence of a planar wall, where only
cylindrical symmetry survives in wave vector space. Moreover, 
in wetting phenomena the correlation length parallel to the surface $\xi_{||}$ 
and the correlation length perpendicular to the surface $\xi_{\perp}$  
are related by \cite{lipowsky,wetting}:
\begin{equation}
\xi_\perp\sim\left\{\begin{array}{ll} \left[\log\left(\xi_\parallel\right)\right]^{\frac{1}{2}} & d=3 \\ \xi_\parallel^{\frac{3-d}{2}} & d<3 \end{array}\right. \label{perptopar}
\end{equation}
This relationship suggests that the fluctuations perpendicular to the wall 
diverge much more slowly than the fluctuations parallel to the wall. 
Therefore it is natural to define a sharp {\em cylindrical} cut-off \cite{Orlandi} 
which prevents long wavelength CW critical fluctuations. Within the HRT approach,
this is implemented by defining the sequence of intermediate potentials: 
\begin{equation}
w^Q(\kappa, q)=\left\{\begin{array}{ll} w(\kappa, q) & \kappa \geq Q \\ 
0 & \kappa<Q \end{array}\right. 
\label{cutoff}
\end{equation}
where $\kappa$ is the component of the wave vector parallel to the surface and $q$ is the component 
normal to the wall. 
The system characterized by the potential $v^Q(\mathbf{r})=v_R(\mathbf{r})+w^Q(\mathbf{r})$ 
will be named $Q$-system. For $Q\to\infty$, $v^Q$ reduces to $v_R$ while for $Q\to0$ the 
full interaction is recovered. Therefore it is natural to look for evolution equations 
governing the change in the physical properties of the $Q$-system as the cut-off is varied. 
In order to derive such evolution equations, we consider the structure of the 
perturbative series which defines the free energy of the model.
In the perturbative diagrammatic expansion of the free energy of the $Q$-system $A^Q$, 
every loop contains one or more $\phi=-\beta w^Q$  bonds \cite{HRT}. Therefore, 
due to the vanishing of $w^Q(\kappa,q)$ for $\kappa < Q$, $A^Q$ is defined by the same
perturbative expansion as the full free energy of the model, $A$, where 
each (longitudinal) momentum integration is limited by an infrered cut-off $Q$.
The correspondence between a cut-off in the potential and a cut-off in fluctuations (i.e. in the 
momentum integrations) is valid with the single exception of the first, mean field, diagram of 
the perturbation series which does not contain any loop. This contribution is in fact 
discontinuous in $Q$ because it contains just the $\kappa=0$ Fourier component of the 
attractive potential (\ref{cutoff}) which is zero for every $Q\neq 0$ and finite for $Q=0$. 
However we can introduce a modified free energy $\mathcal{A}^Q$ which is continuous in $Q$ 
and is simply related to $A^Q$
\begin{equation}
-\beta \mathcal{A}^Q=-\beta A^Q-
\frac{1}{2}\left[\phi(\mathbf{r}=0)-\phi^Q(\mathbf{r}=0)\right] \int d^d\mathbf{r}\;\rho(\mathbf{r})+
\frac{1}{2}\int d^d\mathbf{r}_1d^d\mathbf{r}_2\; \left[
\phi(\mathbf{r}_1,\mathbf{r}_2)-\phi^Q(\mathbf{r}_1,\mathbf{r}_2)\right]
\rho(\mathbf{r}_1)\rho(\mathbf{r}_2) 
\label{modifiedA}
\end{equation}
Analogously, we can introduce the direct correlation functions $C^Q_n$, for each
$Q$-system by  functional derivation of the free energy $A^Q$ with respect to the local density 
$\rho(\mathbf{r})$. 
These correlation functions are continuous in $Q$ for $n\geq 3$ because there are no zero 
loop contributions in the corresponding perturbative expansion, but the two point 
direct correlation function is discontinuous at $\kappa=Q$. So we introduce a
modified function which is continuous, being the second functional 
derivative of $\mathcal{A}^Q$ with respect to the density:
\begin{equation}
\mathcal{C}_{2}^Q(q,\kappa)=C^Q_{2}(q,\kappa)+\left[ \phi(q,\kappa)-\phi^Q(q,\kappa)\right]. 
\label{modifiedC}
\end{equation}
It is apparent from the definitions (\ref{modifiedA}) and (\ref{modifiedC}) that 
$\mathcal{A}^Q$ and $\mathcal{C}^Q_2$ coincide with the free energy and direct correlation 
function of the fully interacting system respectively when $Q\rightarrow 0$. 
On the other hand in the limit $Q\rightarrow\infty$ these modified quantities reproduce the 
mean field approximation for the free energy and the direct correlation function, contrary 
to $A^Q$ and $C_2^Q$ which reduce to the reference system quantities.
This suggests that the HRT procedure does indeed describe the growth of fluctuations on
top of the mean field approximation, as in the RG approach. 

The simple relationship between the cut-off-dependent modified quantities
($\mathcal{A}^Q$ and $\mathcal{C}^Q_2$) and the $Q$-system properties 
($A^Q$ and $C^Q_2$) allows to derive the evolution equations describing how $\mathcal{A}^Q$ and $C_n^Q$ change 
when $Q$ is decreased from infinity to zero, {\em i.e.} when fluctuations on
larger and larger length-scales are included.
The perturbative expansion of the free energy can be specialized to the case
where the reference system is the $Q$-system and the perturbation potential is 
\begin{equation}
  \delta v^Q(q,\kappa)=v^{Q-\delta Q}(q,\kappa)-v^Q(q,\kappa)=
\left\{ \begin{array}{ll} v(q,\kappa) & Q-\delta Q<\kappa<Q \\ 0 & \hbox{elsewhere} \end{array}\right.
\end{equation}
where $\delta Q > 0 $ is an infinitesimal shift in the cut-off. By summing all terms linear in $\delta Q$ in the diagrammatic expansion
of the free energy and taking the limit $\delta Q\rightarrow 0$ one obtains the evolution equation for the free energy. 
Till now we have formally considered the total free energy, but for an inhomogeneous system
it is useful to separate the bulk quantities, which refer to the homogeneous system without an external potential, 
from the surface terms due to the presence of the wall
\begin{eqnarray}
&&\mathcal{A}^Q=\mathcal{A}^Q_b+\Delta\mathcal{A}^Q \nonumber \\
&&\rho(\mathbf{r})=\rho_b+\Delta\rho(\mathbf{r}) \nonumber \\
&&F^Q_2(\mathbf{r}_1,\mathbf{r}_2)=F^Q_{2b}(z_1-z_2,|\mathbf{s}_1-\mathbf{s}_2|)+
\Delta F^Q_2(\mathbf{r}_1,\mathbf{r}_2) \nonumber
\end{eqnarray}
where the two point correlation function $F^Q_2(\mathbf{r}_1,\mathbf{r}_2)$ is the functional 
inverse of $-C^Q_2(\mathbf{r}_1,\mathbf{r}_2)$ {\em i.e.}
\begin{equation}
\int d\mathbf{r}^d\; F^Q_2(\mathbf{r}_2,\mathbf{r}_3)C^Q_2(\mathbf{r}_3,\mathbf{r}_2)=
-\delta(\mathbf{r}_1-\mathbf{r}_2). 
\label{oz} 
\end{equation}
This differs from the known Ornstein-Zernike equation for inhomogeneous fluids because
of the inclusion of the ideal gas terms. 
Note that, due to the cylindrical cut-off, $F_{2b}^Q$ has lost the full rotational 
symmetry of the physical system for every $Q\in (0,\infty)$. 
The separation in bulk and surface terms
may be inserted in the diagrammatic expansion of the total Helmholtz 
free energy. We begin with the derivation of the evolution equation for the 
bulk contribution to the free energy.
Only the one loop diagrams containing 'bulk' $F^Q_{2b}$ bonds alone in the diagrammatic 
expansion contribute to the leading order in $\delta Q$.  
The sum of these term can be carried out in closed form \cite{HRT} and 
reproduces the known random phase formal expression: 
\begin{eqnarray}
\beta \mathcal{A}^Q_b-\beta \mathcal{A}_b^{Q-\delta Q}=-\frac{V}{2}\int \frac{dq}{2\pi}\int_{(Q,Q-\delta Q)} \frac{d^{d-1}\kappa}{(2\pi)^{d-1}}\;\log(1-\phi(q,\kappa)F^Q_{2b}(q,\kappa)).
\end{eqnarray}
This equation is not yet suitable for taking the $\delta Q\rightarrow 0$ limit 
because it contains the discontinuous function $F^Q_{2b}(q,\kappa)=-1/C^Q_{2b}(q,\kappa)$. 
Expressing $F^Q_{2b}$ in terms of the continuous function 
$\mathcal{F}^Q_{2b}(q,\kappa)=-1/\mathcal{C}^Q_{2b}(q,\kappa)$ via Eqs. (\ref{modifiedC}) and  (\ref{oz}) we obtain:
\begin{eqnarray}
\frac{ \partial }{\partial Q} \left( \frac{\beta \mathcal{A}^Q_b}{V}\right)=\frac{1}{2}K_{d-1}Q^{d-2}\int 
\frac{dq}{2\pi}\log(1+\mathcal{F}^Q_{2b}(q,Q)\phi(q,Q)) \label{bulkeveq}
\end{eqnarray} 
where $K_{d-1}$ is a geometric factor defined by 
$$ 
\int \frac{d^{d-1}\kappa}{(2\pi)^{d-1}}\;\delta^{d-1}(\kappa -Q)=K_{d-1} Q^{d-2}. 
$$ 
The initial condition must be imposed at $Q=\infty$ and coincides with the mean field 
approximation for the Helmholtz free energy of the fully interacting system.
Unfortunately, equation (\ref{bulkeveq}) is not written in closed form because the evolution 
of the free energy depends on the knowledge of the two-particle correlation function 
$\mathcal{F}^Q_{2b}(q,Q)$ at a generic value of cut-off $Q$. However, we can study how 
the two-particle function $\mathcal{F}_{2b}^Q$ itself (or equivalently $\mathcal{C}^Q_{2b}$) 
is modified owing to the inclusion of fluctuations, by performing the same analysis 
leading to (\ref{bulkeveq}). In this way we obtain an evolution equation for the pair 
correlation function, but this equation contains the three-particle and the four-particle 
correlation functions. The procedure may be iterated to higher orders thereby obtaining 
an infinite hierarchy of coupled differential equations. This hierarchy may be truncated 
at the level of the first equation by introducing a suitable anstanz for the 
two-particle correlation function and using the compressibility sum rule \cite{HansenMacDonald} 
which relates the second density  derivative of free energy to the zero 
wave-vector limit of the pair correlation function
\begin{equation}
\mathcal{C}^Q_{2b}(\mathbf{k}=0)=
\frac{\partial^2 \left(-\beta\mathcal{A}^Q_b/V\right)}{\partial \rho_b^2} \label{compsumrule}.
\end{equation}
Note that this equation is exact for each $Q$-system, it provides an extremely useful 
link between thermodynamics and the long wavelength limit of correlation functions
but it is not sufficient to close the hierarchy because it does not give information about the momentum 
dependence of $\mathcal{F}^Q_{2b}(q,\kappa)$.

In the derivation of the bulk evolution equation, translational invariance allowed to perform 
the analytical resummation of the one loop diagrams leading to Eq. (\ref{bulkeveq}) without 
specifying the form of $\mathcal{F}^Q_{2b}$. Instead, in order to express in closed form the  
sum of the required diagrams in the expansion of the surface free energies, we need to 
specialize to a particular form of the inhomogeneous two point correlation function 
$\Delta F_2^Q(\mathbf{r}_1,\mathbf{r}_2)$. Therefore, it is important to understand 
what type of fluctuations are described by $\Delta F_2^Q(\mathbf{r}_1,\mathbf{r}_2)$. 
Let us consider the case of a wetting transition between two bulk phases, 
$\beta$ and $\alpha$, in contact with a wall. At a temperature larger than the wetting 
temperature and slightly off coexistence, the system will consist of a slab of the wetting phase, 
with density $\rho_\beta$, an interfacial region and the bulk $\rho_\alpha$ phase. 
If $F^{\alpha}_{2b}$, which is the bulk correlation function of the $\alpha$ phase, 
is not equal to $F^{\beta}_{2b}$ then  $\Delta F_2^Q(\mathbf{r}_1,\mathbf{r}_2)$, 
which is defined as $F_2(\mathbf{r}_1,\mathbf{r}_2)-F_{2b}(\mathbf{r}_1-\mathbf{r}_2)$, 
describes not only the correlations in the interfacial region but also in the slab. 
In this case it is difficult to find a suitable ansatz for the form of $\Delta F_2^Q$. 
However, in systems where the bulk correlation functions are the same for the two phases
the problem simplifies and $\Delta F_2^Q$ just describes the interfacial region. 
This is precisely the case of the lattice gas model with nearest-neighbor interaction 
(i.e. of the Ising model) due to the symmetry $\rho_\beta=1-\rho_\alpha$. If the inhomogeneous
part of the two point correlation function describes only the interfacial region we can 
parametrize $\Delta \mathcal{F}_2^Q$ according to the standard factorized form \cite{Henderson}:
\begin{equation}
\Delta \mathcal{F}^Q_2(q_1,q_2,\kappa)=g^Q(q_1)g^Q(q_2)F_s^Q(\kappa) \label{coras}
\end{equation}
where $g^Q(z)$, the Fourier transform of $g^Q(q)$, is usually identified as the 
first spatial derivative of the 
density profile and we have taken the wall perpendicular to $z$ axis. 
Using this parametrization it is possible to perform the summation of the one loop diagrams 
which allows to obtain the evolution equation for the surface free energy:
\begin{eqnarray}
\frac{\partial}{\partial Q}\left( \frac{\beta\Delta\mathcal{A}^Q}{S}\right)=
\frac{K_{d-1}}{2}Q^{d-2}\log(1+F^Q_s(Q)\alpha^Q(Q)) 
\label{sureq}
\end{eqnarray}
where $\alpha^Q$ is:
\begin{equation}
\alpha^Q(\kappa)=\int \frac{dq}{2\pi}\;g^Q(q)
\frac{\phi(\kappa,q)}{1+\phi(\kappa,q)\mathcal{F}^Q_{2b}(\kappa,q)}g^Q(-q). 
\label{alphaA}
\end{equation}
Note that the evolution of the surface free energy is coupled to the bulk 
evolution through the $\alpha^Q$ coefficient. 

In the general case of $F^{\alpha}_{2b}\neq F^{\beta}_{2b}$ 
the derivation of the evolution equation is more involved: 
By introducing further simplifying assumptions in the spirit of the local 
density  approximation, it is still possible to obtain a decoupled evolution equation 
identical to (\ref{sureq}) with a slightly different coefficient $\alpha^Q$ \cite{phd}:  
\begin{equation}
\alpha^Q_{LDA}(\kappa)=\int \frac{dq}{2\pi}\;g^Q(q)\phi(\kappa,q)g^Q(-q). 
\label{LDAalphaA}
\end{equation}
Equation (\ref{sureq}) has been obtained by starting from an expansion of the free energy 
at constant density profile: every $Q$-systems has the same $\rho(\mathbf{r})$. It means that
the external potential which stabilizes the density profile changes with $Q$.  
In the case of wetting or drying this procedure is clearly artificial: we would rather 
want to study how the density profile is modified by the inclusion of fluctuations 
at {\em fixed} external potential. In order to allow the change in $\rho(\mathbf{r})$ 
when fluctuations are included, it is convenient to perform a Legendre transform 
of the total free energy $\mathcal{A}^Q$. In this way, it is possible to obtain the 
evolution equations at fixed fugacity, $i.e.$ by keeping fixed the quantity 
$$
\gamma(z)=\beta\left[\mu-U(z)\right],
$$ 
where $U(z)$ is the microscopic interaction between the wall and the molecules of the fluid. 
We first define the (modified) grand free energy by
\begin{equation}
-\beta\mathcal\omega^Q=-\beta\mathcal{A}^Q+\int d^3\mathbf{r}\; \rho^Q(z)\gamma(z)
\end{equation}
Here $\gamma(z)$ is fixed to the physical value, while the density profile 
changes with the cut-off wave vector $Q$. The  $Q$-dependent density profile 
is related to $\omega^Q$ by:
\begin{equation}
\rho^Q(z)=-\frac{\delta \beta\omega^Q }{\delta \gamma(z)}.
\end{equation}
As usual, for $Q\rightarrow \infty $ $\omega^Q$ reduces to the mean field grand
free energy and $\rho^Q(z)$ approaches the density profile in mean field approximation. 
From the properties of the Legendre transform we obtain:
\begin{equation}
\left( \frac{\partial \beta\omega^Q}{\partial Q}\right)_\gamma =
\left( \frac{ \partial \beta\mathcal{A}^Q}{\partial Q}\right)_\rho
\end{equation}
As a consequence, the evolution equation for the grand free energy formally coincides 
with that of the Helmholtz free energy and the surface tension
$\sigma^Q=\Delta \omega^Q/S$ obeys the evolution equation: 
\begin{equation} 
\frac{\partial (\beta \sigma^Q)}{\partial Q}=
\frac{1}{2}K_{d-1}Q^{d-2}\log(1+F^Q_s(Q)\alpha^Q(Q)). 
\label{surfteneq}
\end{equation}
Note that now $\alpha^Q$ is given again by Eq. (\ref{alphaA}) 
in terms of $g^Q(q)$ and $F_s^Q(\kappa)$. Analogously to the bulk case, 
also surface quantities obey a hierarchy of coupled differential equations which 
can be closed at the free energy level (\ref{surfteneq}) by a suitable 
ansatz for $F^Q_s(\kappa)$ and $g^Q(z)$. Two surface sum rules 
will provide the link between the free energy and the correlation functions \cite{wetting}:
\begin{eqnarray}
&&-\frac{\partial \sigma^Q}{\partial \mu}=
\Gamma^Q=\int dz\;\left[\rho^Q(z)-\rho^Q_b\right]=\int dz\;\Delta\rho^Q(z) 
\label{assorbsumrule}\\
&&-\frac{\partial^2 \sigma^Q}{\partial \mu^2}=
\int dz_1dz_2\;\Delta\mathcal{F}^Q_2(z_1,z_2,\kappa=0)=\beta F^Q_s(0)\left|\int dz\; g^Q(z)\right|^2. 
\label{surfcompsumrule}
\end{eqnarray}
Equation (\ref{surfcompsumrule}) can be identified as a ``surface compressibility'' sum rule
while (\ref{assorbsumrule}) constrains the form of the density profile. It is interesting to note that 
this two sum rules are sufficient to obtain a quite good description 
of the wetting phase transition \cite{Henderson}.

\subsection{Asymptotic Equations}
\subsubsection{Bulk Equation}
In this section we show how the HRT evolution equations describe the 
asymptotic regime near phase transitions. 
The standard bulk HRT with spherical cut-off in the asymptotic critical region 
is known to give the same evolution 
equation obtained by momentum space RG approach \cite{HRT,HRT1}.  Here we show that the introduction of a 
cylindrical cut-off does not modify this property of the bulk HRT equation.

In the thermodynamic states close to the bulk critical point, equation (\ref{bulkeveq}) can be simplified. 
This regime is characterized by the growth of long-range fluctuations, that is by the divergence of 
$\mathcal{F}^Q_{2b}(q=0,\kappa=0)$ for $Q\rightarrow 0$ {\em i.e.} in the final stages of the evolution. 
In order to study the critical region we have to extract the singular contribution to the free energy from 
the evolution equation (\ref{bulkeveq}): Only a small neighborhood of the integration domain, close to $q=0$ and $Q\sim 0 $ 
contributes to the singularity. In this domain we can simplify the argument  of the logarithm due to the divergence 
of the pair correlation function for $q$ and $\kappa$ approaching zero. 
\begin{eqnarray}
\frac{\partial }{\partial Q}\left( \frac{\beta \mathcal{A}_b^Q}{V}\right) =
\frac{K_{d-1}}{2}Q^{d-2}\int_{-q_0}^{q_0}\frac{dq}{2\pi}\;\log\left[\mathcal{F}^Q_{2b}(Q,q)\right] \nonumber
\end{eqnarray}
Here $q_0$ is an arbitrary ultraviolet cut-off which is introduced to extract the singular contribution 
to the integral. 
Note that this equation does not contain the inter-atomic potential explicitly and acquires a universal form 
independent of the specific microscopic interaction. The next step is to choose a form for the bulk pair 
correlation function. As usual, we adopt an Ornstein-Zernike form for the direct correlation function 
\begin{equation}
\mathcal{C}^Q_{2b}(\kappa,q)=-\frac{1}{\mathcal{F}_{2b}^Q(\kappa,q)}=
-b(\kappa^2+q^2)+\frac{\partial^2}{\partial \rho^2}\left( \frac{-\beta \mathcal{A}_b^Q}{V}\right) \nonumber
\end{equation}
where we imposed the sum rule (\ref{compsumrule}) and we assumed that $b$ remains finite also at the critical point. 
This particular ansatz for the pair correlation function implies that the critical exponent $\eta$ is zero. 
We know that in three dimension $\eta$ is not zero, but its value is small and it is zero to first order in the  
$\epsilon$-expansion \cite{Wilson} so that, in three dimension, this is a good approximation. 
Substituting this parametrization in the evolution equation we obtain
\begin{eqnarray}
\frac{\partial}{\partial Q}\left(\frac{-\beta \mathcal{A}_b^Q}{V}\right)=
\frac{K_{d-1}}{2}Q^{d-1}\int^{+\frac{q_0}{Q}}_{-\frac{q_0}{Q}}\frac{du}{2\pi}\;
\log\left[Q^2b(1+u^2)-\frac{\partial^2}{\partial \rho^2}\left( \frac{-\beta \mathcal{A}_b^Q}{V}\right)\right]\nonumber
\end{eqnarray} 
We can eliminate the explicit dependence on the cut-off $Q$ in the evolution equation by another change of variable
\begin{eqnarray}
&&t=-\log(Q) \nonumber \\
&&z= \left( \rho_b-\rho_{bc}\right)\sqrt{\frac{b}{K_{d-1}}} e^{\frac{2-d}{2}t} \nonumber \\
&&H_t(z)=-\frac{\beta}{V}\left( \mathcal{A}_{bc}^Q-\mathcal{A}_b^Q\right)\frac{e^{dt}}{K_{d-1}} \nonumber 
\end{eqnarray}
where $\rho_{bc}$ and $\mathcal{A}_{bc}$ are the density and the free energy at the critical point. Using these variables
the asymptotic evolution equation becomes:
\begin{eqnarray}
\frac{\partial H_t(z)}{\partial t}=dH_t(z)+\frac{2-d}{2}zH'_t(z)+\frac{1}{2}\int^{+\infty}_{-\infty}\frac{du}{2\pi}\;\log\left[\frac{H''(z)+u^2+1}{H''(0)+u^2+1}\right]
\end{eqnarray}
where $H'(z)$ and $H''(z)$ are the first and the second derivative of $H$ with respect to $z$ and we set $q_0/Q\rightarrow \infty$ in the $Q\rightarrow 0$ limit. The integration can be easily
performed and we obtain:
\begin{eqnarray}
\frac{\partial H_t(z)}{\partial t}=dH_t(z)+\frac{2-d}{2}zH'_t(z)+
\frac{1}{2}\left[\sqrt{H''_t(z)+1}-\sqrt{H''_t(0)+1}\right]. 
\label{asybulkeveq}
\end{eqnarray}
This equation can be analytically investigated in $d=4-\epsilon$ dimensions. 
The fixed point and the relevant eigenvalues provide the critical exponents to first order in $\epsilon$:
$\gamma=1+\epsilon/6$, $\beta=1/2-\epsilon/6$ and 
$\delta=3+\epsilon$ which coincide with the known exact RG results \cite{Wilson}.
As expected, to lowest order in $\epsilon$, the choice of the cut-off symmetry does not modify 
the universal properties close to the bulk critical point within the HRT approach.

\subsubsection{Surface Equation}
The surface evolution equation we have obtained (\ref{sureq}) depends on the form of the 
density profile through the quantity $\alpha^Q(\kappa)$ (\ref{alphaA},\ref{LDAalphaA})
if $g^Q(z)$ is taken to represent the spatial derivative of $\rho^Q(z)$. 
Here and in the following we assume 
that the shape of $\rho^Q(z)$ is not affected by fluctuations. Fluctuations are assumed to 
shift rigidly the mean field density profile by a $Q$-dependent amount $l_Q$ identified as 
the distance of the interface from the wall: $\Delta \rho^Q(z)= \Delta \rho f(z-l_Q)$. 
This assumption can be related to the non-homogeneity of the spectrum of fluctuations 
and is customary in most studies of the wetting phenomena \cite{fisher,lipowsky}.
From Eq.(\ref{perptopar}) we see that the fluctuations perpendicular to the interface, 
{\em i.e.} the fluctuations which distort the form of the profile, are only weakly divergent 
and as a first approximation we can neglect their effect. Note that this assumption is not justified 
near the bulk critical point, where the interplay between bulk and surface fluctuations does modify 
also the form of the density profile. 
Our analysis is therefore valid only away from the bulk critical point. 
The long wavelength behavior of the surface correlation function $F^Q_s(\kappa)$ is 
modeled by an Ornstein-Zernike form: 
\begin{equation}
 F^Q_s(\kappa)=\left[ -(\Delta\rho)^2\left(\frac{\partial^2 \beta \sigma^Q}
{\partial \left(\beta\mu\right)^2}\right)^{-1}+b_s\kappa^2\right]^{-1} 
\end{equation}
where $b_s$ is a non-universal constant that tends to a finite limit at the wetting transition 
and we have used the exact sum rule (\ref{surfcompsumrule}):
\begin{equation}
\frac{\partial^2 \left(-\beta\sigma^Q\right)}{\partial \left( \beta\mu\right)^2}=(\Delta\rho)^2F^Q_s(\kappa=0).
\end{equation} 
where $\Delta\rho=\int dz g^Q(z)=(\rho_\alpha-\rho_\beta)$ is the difference between 
the bulk density of the two phases across the interface. Because we assume to be away from bulk critical point this quantity is regular
at wetting transition. The assumptions we made on the density profile and on the surface correlation function are similar to those
underlying the usual RG group treatment of the wetting transition \cite{lipowsky,lipowsky1,fisher,wetting}. 
Note that for wetting phenomena the critical exponent $\eta$ is known to be zero \cite{wetting,Forgacs} 
but the Ornstein-Zernike form for the surface correlation function is strictly accurate 
only in the interfacial region \cite{Parry2,Parry3}.

Away from the bulk critical point, the asymptotic form of the evolution equation for the 
surface free energy then becomes:
\begin{equation}
\frac{\partial (-\beta \sigma^Q)}{\partial Q}=
\frac{K_{d-1}}{2}Q^{d-2}\log\left(\left [  \frac{\partial^2(-\beta\sigma^Q)}
{\partial (\beta\mu)^2}\right ]^{-1}+b'_sQ^2\right)
\label{RGseq}
\end{equation}
where $b'_s=b_s(\Delta\rho)^2$ and we have disregarded non singular terms.
Notice that eq.(\ref{RGseq}) does not depend on the specific shape of the density profile. 
The asymptotic evolution equation (\ref{RGseq}) 
is in fact fully equivalent to the non-perturbative ``functional" RG approach \cite{lipowsky,lipowsky1}. 
To prove this remarkable correspondence, we introduce the surface free energy $a^Q(\Gamma)$ by
Legendre transform of $\sigma^Q(\mu)$:
\begin{equation}
a^Q(\Gamma)=\sigma^Q(\mu^Q)+\mu^Q\Gamma \, .
\end{equation}
Here $\Gamma$ is the adsorption (\ref{assorbsumrule}) 
and then $\mu^Q$ is such that
\begin{equation}
\mu^Q=\frac{\partial a^Q(\Gamma)}{\partial \Gamma}.
\end{equation}

The asymptotic evolution equation of $a^Q(\Gamma)$ follows straightforwardly from 
the evolution equation of $\sigma^Q$ (\ref{RGseq}) and from the properties of the Legendre transform 
\begin{equation}
\frac{\partial (-\beta a^Q)}{\partial Q}=\frac{K_{d-1}}{2}Q^{d-2}
\log\left(-\frac{\partial^2(-\beta a^Q)}{\partial \Gamma^2}+b_sQ^2\right).
\end{equation}
By means of the change of variables:
\begin{eqnarray}
&&t=-\log (Q) \nonumber \\
&&l=\Gamma\left(\frac{b'_s}{K_{d-1}}\right)^{{1\over2}}e^{{(d-3)\over 2}t}\nonumber \\
&&V_t(l)=\frac{\beta e^{(d-1)t}}{K_{d-1}} a^Q\nonumber 
\end{eqnarray}
we obtain the known RG equation \cite{lipowsky,lipowsky1,fisher,Forgacs}:
\begin{equation}
\frac{dV_t(l)}{dt}=(d-1)V_t(l)+\frac{1}{2}(3-d)
\frac{\partial V_t(l)}{\partial l}+\frac{1}{2}
\log\left[\frac{\partial^2V_t(l)}{\partial l^2}+1\right]. 
\label{Veveq}
\end{equation}
where $V_t(l)$ is the effective interfacial potential in the RG approach. From this formal 
equivalence, we can identify the renormalized effective interaction $V_t(l)$ as the 
surface Helmholtz free energy $a^Q$ and the height of the interface $l$ as a measure 
of the absorption $\Gamma$. The initial ($i.e.$ bare) form of $V_t(l)$ is then given by 
the value of $a^Q(\Gamma)$ at a suitable matching cut-off $Q=Q_0\ll1$. 
The appropriate value should include the effects of short wavelength fluctuations 
($i.e.$ of fluctuations at $Q> Q_0$) but a rough estimate is obtained starting 
from the mean field free energy. It is possible to show that, in the sharp-kink 
approximation \cite{wetting} for the density profile of a continuous system, 
the initial form of our $V_t(l)$ is given by (\ref{efpotential})
{\em i.e.} the same used in the effective capillary wave Hamiltonian.

\section{Lattice gas complete wetting}
\subsection{Mean field}
In this section we consider the wetting transition in a semi-infinite lattice 
gas model with nearest-neighbor interaction
 in contact with an unstructured wall. The interaction between the wall 
and the lattice gas is described by an external potential $u_i$, where the 
subscript $i \geq 1$ labels layers of the lattice gas: we consider a planar geometry 
with an external potential translationally uniform in directions parallel to the wall. 
It is convenient to introduce an interlayer interactions $v_{ij}$ which collects all 
inter-atomic nearest-neighbor interactions of the particles belonging to the $i$-th 
layer and the $j$-th layer, which are assumed to be nearest neighbors.
Note that there is also a contribution $v_{ii}$ which describes the interaction 
of particles whitin the same layer. We shall discuss in the following  only the 
particular model for which the external potential acts only on the first layer and is attractive
$$ u_i=u \delta_{i,1} $$
and
$$ v_{ij}=\left\{\begin{array}{lll} v & i,j\hbox{ nearest-neighbors layers} \\ 
4v & i=j \\0 & \hbox{otherwise.} \end{array}\right. $$
with $u,v<0$. In mean-field approximation, the free energy $A_{mf}$ of the system reads
\begin{equation} 
\frac{-\beta A_{mf}}{S}=-\frac{\beta A_R}{S}+\frac{1}{2}\sum_{i,j}\rho_i\rho_j\phi_{ij} 
\end{equation}
where $\rho_i$ is the density of the $i$ layer, $\phi_{ij}=-\beta v_{ij}$, $S$ is the area of 
the layers. The reference system is a hard core lattice gas. 
\begin{equation}
\frac{-\beta A_R}{S}=-\sum_i\left[(1-\rho_i)
\log\left(1-\rho_i\right)+\rho_i\log(\rho_i)\right].
\end{equation}

It is well know that in mean-field approximation the bulk critical temperature is 
$k_BT_c/|v|=3/2$. For $T<T_c$ there are two stable phases, 
$\alpha$ and $\beta$, which coexist for $\mu_0=-3|v|$. This value of the 
chemical potential at coexistence is not modified by fluctuations due to
the particle-hole symmetry of the model, which also leads to a relationship
between the bulk densities of the two coexisting phases: $\rho_{\alpha}=1-\rho_{\beta}$.
The mean field density profile is related to the external potential by thermodynamics:
\begin{equation}
\gamma_i=-\frac{\partial}{\partial \rho_i}\left(\frac{\beta A_{mf}}{S}\right)=
\log\left[\frac{\rho_i}{1-\rho_i}\right]-\sum_j\rho_j\phi_{ij} 
\label{densityequation}
\end{equation}
where $\gamma_i=\beta(\mu-u_i)$.  This is an non linear equation for the density profile
which can be conveniently written as:
\begin{equation}
\rho_i=\frac{e^{x_i}}{1+e^{x_i}}  \label{densityequation1}
\end{equation}
where
$$ x_i=\gamma_i+\sum_j\rho_j\phi_{ji}. $$ \\

\begin{figure}[ht]
\includegraphics[scale=0.45]{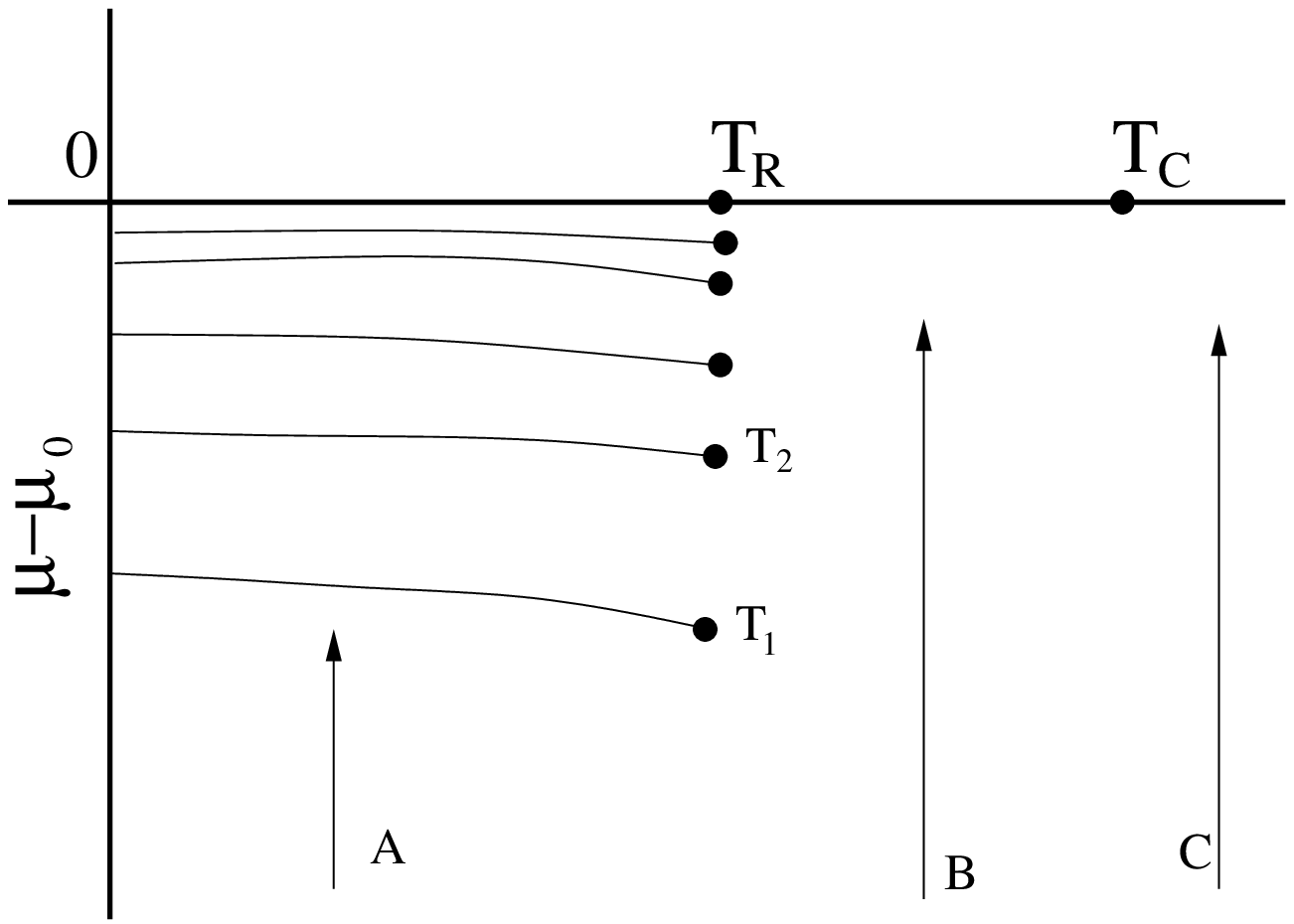}
\phantom{aaa}
\includegraphics[scale=0.45]{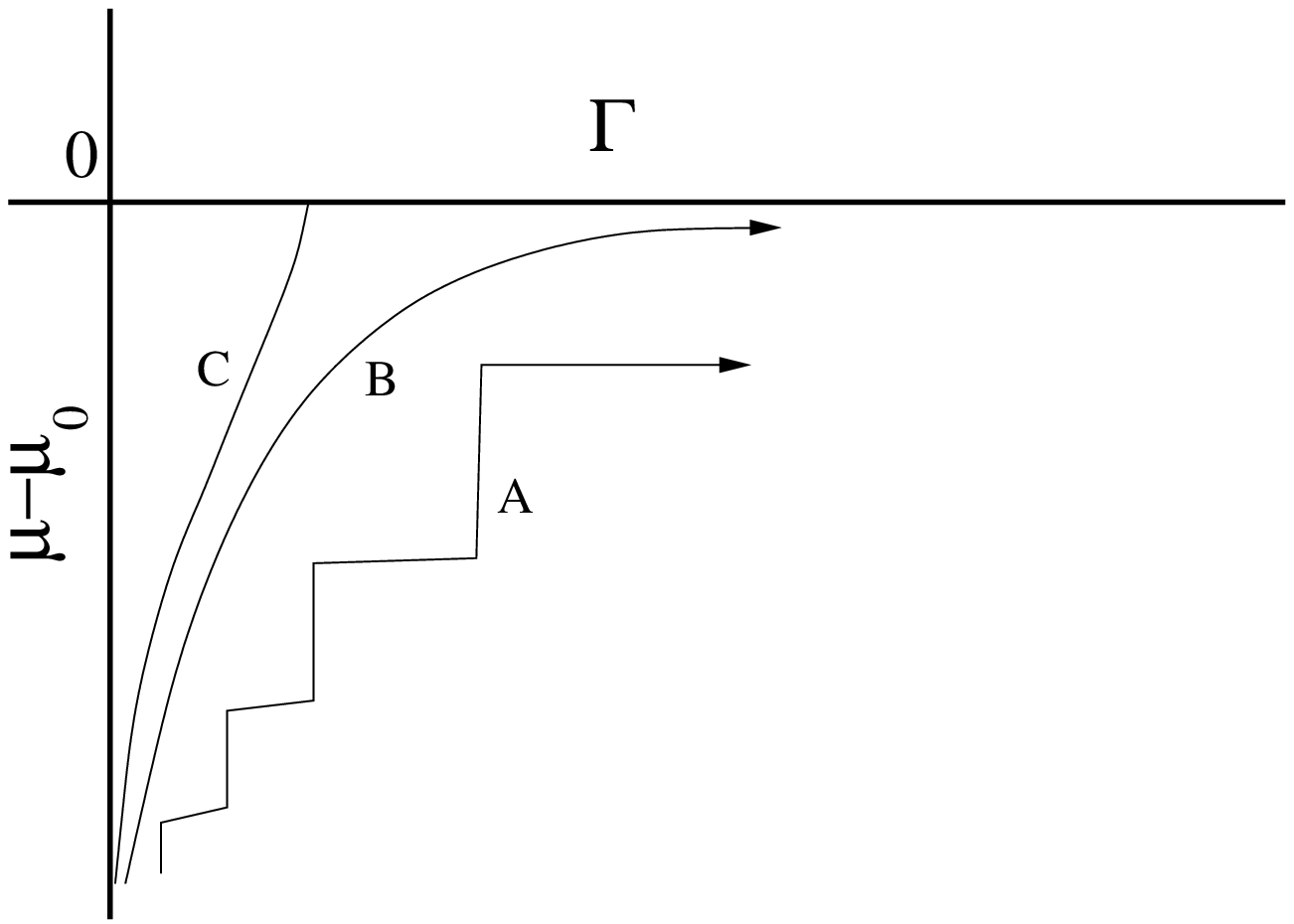}
\caption{\small Tipical surface phase diagram with representative gas-phase adsorption isotherms 
for strong substrate regime ($u/v>1$).$\mu$ is in units $v$ and $\Gamma$ is in units of lattice spacing.}
\label{diagrammastrong}
\end{figure}

\begin{figure}[ht]
\includegraphics[scale=0.45]{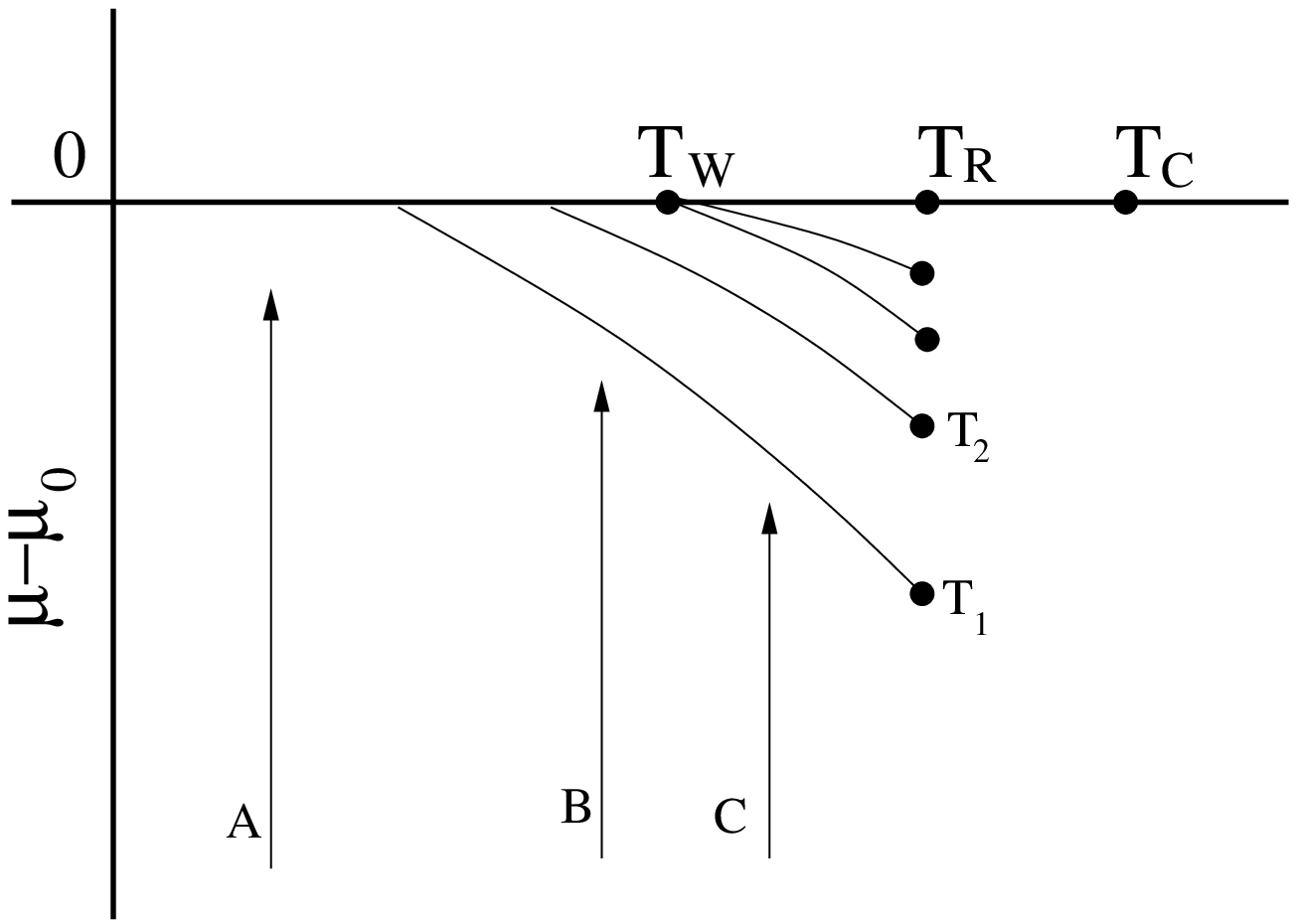}
\phantom{aaa}
\includegraphics[scale=0.45]{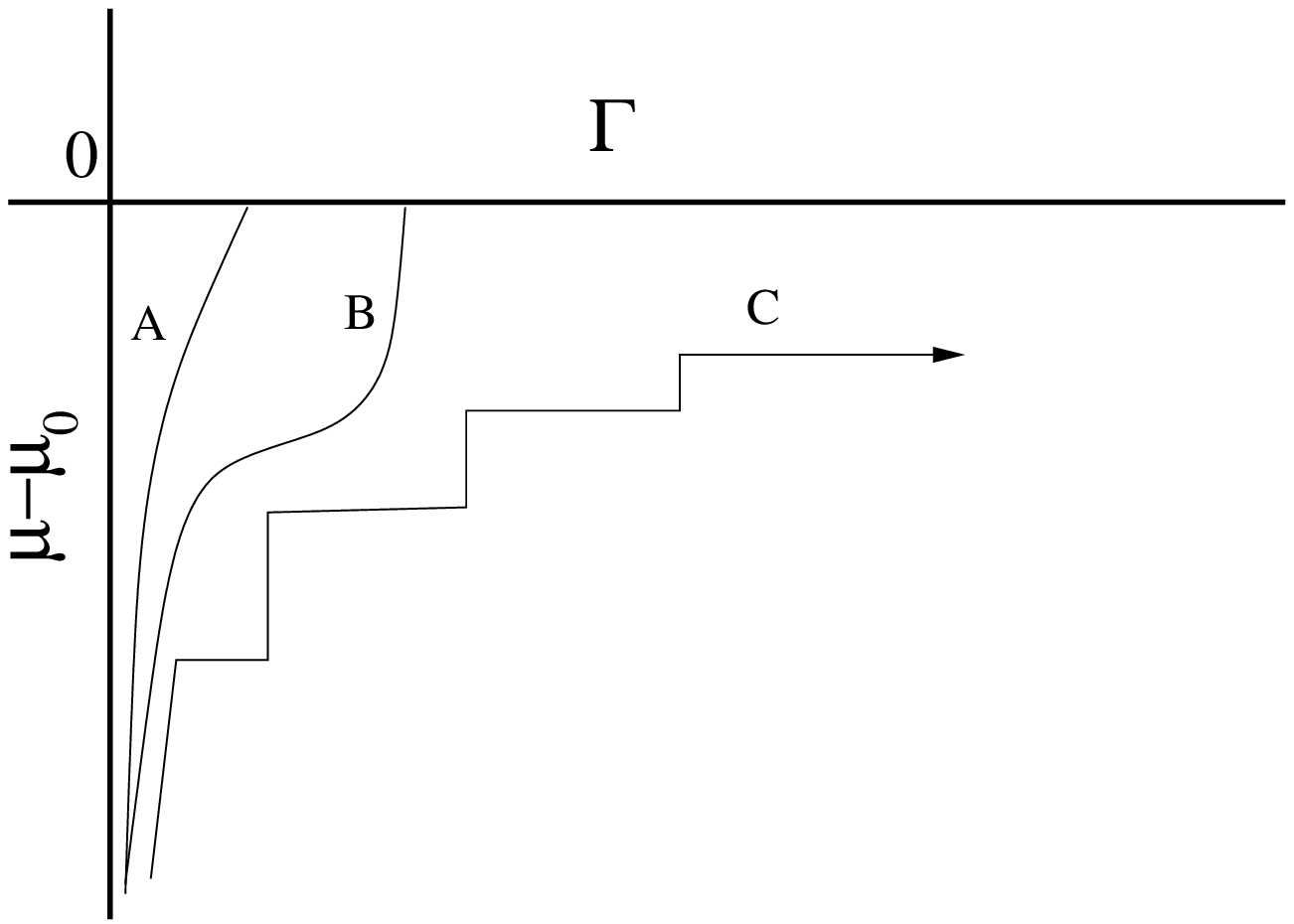}
\caption{\small Tipical surface phase diagram with representative gas-phase adsorption isotherms 
for intermediate substrate regime $1/2<u/v<1$.$\mu$ is in units $v$ and $\Gamma$ is in units of lattice spacing.}
\label{diagrammaintermediate}
\end{figure}

On general grounds, for nearest-neighbor interaction there are two possible scenarios depending on the value of $u/v$ \cite{latticegas1}
named the strong substrate regime ($u/v>1$) and the intermediate substrate regime ($1/2<u/v<1$).
There is also another regime for $0<u/v<1/2$, the weak substrate regime, but its properties follow from those of 
the intermediate regime due to the symmetry of the Ising model: they describe the same phenomena of the intermediate 
regime but on ``the other side'' of the coexistence line \cite{latticegas1}, {\em i.e.} in one case the bulk density 
is $\rho_\alpha$ in  the other case  is $\rho_\beta$.  
In Fig.\ref{diagrammastrong} and Fig.\ref{diagrammaintermediate} we sketch the phase diagrams in the strong and 
intermediate regimes respectively. For $u/v>1$ an infinite sequence of  transitions occurs, corresponding 
to condensation of successive mono-layers. The critical temperatures of these layer transitions approach, for high-oder 
multi-layers, a well defined temperature $T_R$, the roughening temperature. At temperature below $T_R$ isotherms 
for the adsorption $\Gamma$ show an infinite sequence of sharp steps as $\mu\rightarrow\mu_0$. Between $T_R$ and 
$T_c$ the steps are rounded but $\Gamma$ still diverges approaching coexistence. Instead for $u/v<1$ the layer transitions 
no longer extend to the $T=0$ axis but now meet the coexistence axis below a characteristic wetting temperature $T_W$.
Notice that the phase diagrams of the intermediate and the strong substrate regimes are similar for $T>T_W$.\\

\begin{figure}[ht]
\includegraphics[scale=0.8,angle=270]{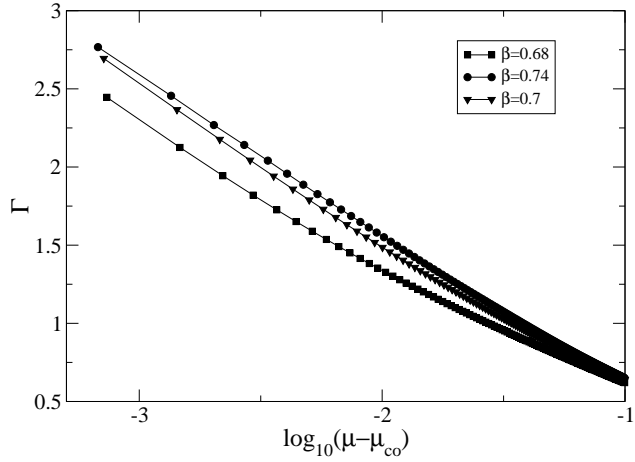}
\caption{\small The logarithm plot of the adsorption isotherm in mean-field approximation for $\beta=v/k_bT=0.74$, 
$\beta=0.70$, $\beta=0.68$ and $u/v=0.95$. We see that in the asymptotic region the three adsorption isotherms 
are well approximated by straight lines. $\mu$ is in units of $v$ and $\Gamma$ is  in units of lattice spacing. }
\label{assorbimentolog}
\end{figure}

Mean-field theory fails to describe the roughening transition and it  incorrectly predicts 
$T_R=T_c$ \cite{latticegas2}. 
\begin{figure}[ht]
\includegraphics[scale=0.8,angle=270]{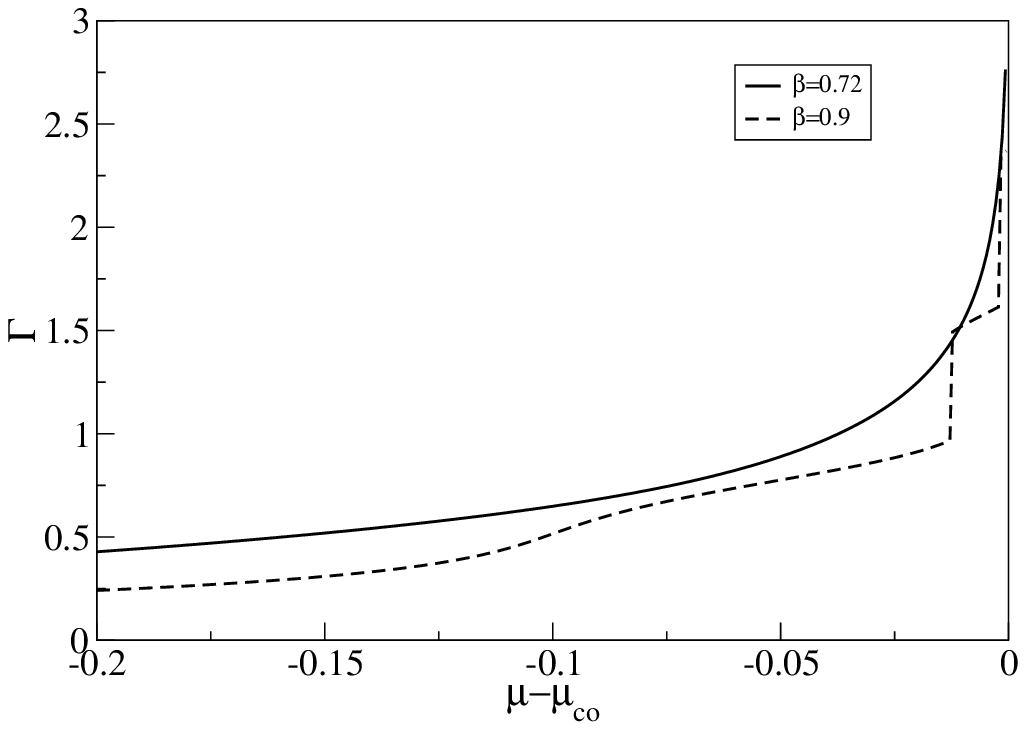}
\caption{\small Two absorption isotherm in the mean-field approximation for $\beta=v/k_bT=0.72$ and 
$\beta=0.9$ ($\beta_c=2/3$) and $u/v=0.95$. We see that for low temperature the layer transitions 
are evident, but for high temperature the isotherm is smooth. $\mu$ is in units $v$ and $\Gamma$ is in units of lattice spacing.}
\label{assorbimento}
\end{figure}
Therefore, at mean-field level there is no smooth wetting transition but only a sequence of layer transitions,  
nevertheless it is interesting to note that in the intermediate regime for high temperatures the behavior of $\Gamma (\mu)$
at fixed $T$ is very similar to that of a smooth wetting transition. 
We numerical solve the non linear mean field density profile equation (\ref{densityequation1}), with standard 
technique \cite{Oliveira}, in the intermediate regime, $u/v=0.95$, at different temperatures 
and we calculate the adsorption
$$ \Gamma=\sum_i (\rho_i-\rho_b) $$
where $\rho_b$ is the bulk density obtained by 
$$\beta\mu=\log\left[\frac{\rho_b}{1-\rho_b}\right]-6\rho_b\phi. $$
At low temperature, as previously discussed, we see a sequence of layer transitions while at high temperature 
the layer transitions are still present but are shifted very near the coexistence line.
As a consequence, the $\Gamma (\mu)$ isotherm, shown in Fig.\ref{assorbimento}, is smooth and looks very similar 
to what is expected in the wetting regime: The divergence close to the coexistence line is well represented by 
$\Gamma(\mu)\simeq A \log\big(\mu-\mu_{co}\big)$ as in wetting. A fit of the data points gives a value $A\simeq 0.45$
in a wide range of temperatures (see Fig.\ref{assorbimentolog}). 
If we suppose that the Capillary Wave effective Hamiltonian describes the critical behavior of wetting phenomena, 
we can use our data on the critical amplitude $A$ to estimate the wetting parameter $\omega$:
\begin{equation}
\left(1+\frac{\omega}{2}\right)=\frac{A}{\Delta\rho \xi_b} 
\label{Aomega}
\end{equation}
where $\Delta\rho$ is the difference between the bulk density of the two phases across the interface and we use $l\simeq \Gamma (\Delta\rho)^{-1}$.
Near coexistence  $\Delta\rho=1-2\rho_b$ due to the particle-hole symmetry of the system. 
The {\em true} correlation length $\xi_b$ \cite{Tarko} can be obtained from the Ornstein-Zernike form 
of the bulk correlation we have adopted in this investigation:
\begin{eqnarray}
\cosh(\xi_b^{-1})&=&1+\frac{1}{2}\left(\xi_b^*\right)^{-2} \\
\left(\xi_b^*\right)^2&=&\frac{\rho_b(1-\rho_b)\beta}{1-6\beta\rho_b(1-\rho_b)}
\label{corlength}
\end{eqnarray}
where $\xi_b^*$ is the second-moment correlation length, analytically obtained by expanding to the second order 
the Fourier transform of the direct correlation function. By substituting into eq. (\ref{Aomega}) our numerical result for critical amplitude $A$, we obtain $\omega_{mf}\simeq 0$ as expected in mean field approximation. \\

We can also evaluate the mean field surface susceptibility $\chi_s$ defined as:
$$\chi_s=\frac{\partial \Gamma}{\partial \mu}=\sum_i\left(\frac{\partial \rho_i}{\partial \mu}-
\frac{\partial \rho_b}{\partial \mu}\right).$$
To obtain $\chi_s$ in mean field, we differentiate the density profile equation (\ref{densityequation}) with respect to 
the chemical potential
$$\beta=\frac{\frac{\partial \rho_i}{\partial \mu}}{\rho_i\big(1-\rho_i\big)}-
4\beta\frac{\partial \rho_i}{\partial \mu}-\beta\frac{\partial \rho_{i+1}}{\partial \mu}-
\beta\frac{\partial \rho_{i-1}}{\partial \mu}. $$
This is a tridiagonal set of linear equations in $\frac{\partial \rho_i}{\partial \mu}$ that can be numerically solved.
This route gives more accurate results than the direct numerical differentiation of the adsorption.
Close to coexistence, the surface susceptibility diverges as $\chi_s\simeq A /\Delta\mu$.

\subsection{HRT surface equation}
After having discussed the mean field approximation, which defines the initial condition
of the HRT equation, 
we are ready to tackle the numerical solution of the HRT surface equation. 
In this first application of the HRT approach to inhomogeneous systems we are mainly 
interested in surface fluctuations in a temperature range not too close to the bulk critical point.
In such a regime, the dependence of the surface evolution equation (\ref{surfteneq})
on bulk quantities via the definition (\ref{alphaA}) can be neglected. Therefore, 
as first step, we have chosen to adopt a local density approximation for $\alpha^Q(\kappa)$ 
(\ref{LDAalphaA}):
\begin{equation}
\alpha^Q(\kappa)=\int \frac{dq}{2\pi}\;\left|g^Q(q)\right|^2\phi(q,\kappa).
\label{alpha1}
\end{equation}
In Appendix A we show how the evolution equation for the surface tension (\ref{sureq}), constructed
for a continuum system, must be modified to be applied to a lattice system  \cite{HRT,HRTlattice}.
The evolution equation for the surface tension of the lattice gas in three dimensions is given by 
\begin{equation}
\frac{\partial \left(-\beta\sigma^Q\right)}{\partial Q}=\frac{1}{2}D_2(Q)\log\left[1+F_s(Q)\alpha(Q)\right] 
\label{latticeeq}
\end{equation}
with $Q\in [-1:1]$ and $D_2(Q)$ is the two dimensional density of states (see Appendix A).

According to the discussion in Section II-A, the factor $g^Q(q)$ in equation (\ref{alpha1}) is 
related to the Fourier transform of the first derivative of the density profile. 
This definition, however, cannot be stricltly carried over to lattice models where the density 
profile is defined only on lattice sites. Instead, we parametrized $g^Q(z)$ in terms
of a rigidly shifted mean field density profile: 
\begin{equation}
g^Q(z)=\rho'_Q(x)=\rho'_{mf}(z-l_Q)
\label{gq}
\end{equation}
where $l_Q$ is the (unknown) $Q$-dependent position of the interface and $\rho_{mf}(z)$ 
is an analytic
interpolation of the mean field $\rho_i$ obtained by fitting the discrete density profile
with a suitable  continuum density profile:
\begin{equation}
\rho_{mf}(z)=\frac{1}{2}\left((a+b)+(a-b)\, {\rm erf}\left(\frac{z}{\xi_\perp}\right)\right). 
\label{mfr}
\end{equation}
This form does adequately represent the actual density profile by taking $a$ equal to the bulk density and 
$b$ to the contact density $\rho_1$. The width $\xi_\perp$ of the interface is only weakly dependent on the
thermodynamic state and diverges logarithmically at wetting. As a first approximation we set $\xi_\perp$
independent of $\mu$. Instead, the position of the interface $l_Q$ does depend on the state and will
be strongly renormalized by fluctuations. In fact, $l_Q$ is easily related to the adsorption $\Gamma_Q$,
via the parametrization (\ref{gq}):
\begin{equation}
\Gamma_Q=\frac{\partial\left(-\beta\sigma^Q\right)}{\partial \beta\mu }=\int_0^\infty dz\;\big(\rho_{mf}(z-l_Q)-\rho_b\big)\simeq
\frac{l_Q}{2}(\rho_1-\rho_b)
\label{sumgammaq}
\end{equation}
Note that, due to the parametrization (\ref{gq}), the value of
$l_Q$ drops out of the definition of $\alpha^Q(\kappa)$ (\ref{alpha1}) which then becomes $Q$-independent
(see Appendix B). We parametrized the  surface correlation function by an OZ form:
\begin{equation}
F^Q_s(\kappa)=\frac{\rho_b\big(1-\rho_b\big)}{\left|\int dz\;\rho'_Q(z)\right|^2\left(1-\lambda_Q\alpha^Q(\kappa)\right)}
\label{surfun}
\end{equation}
in terms of the unkown parameter $\lambda_Q$. When this expression is substituted into Eq.
(\ref{coras}) we get
\begin{equation}
\Delta F^Q(z_1,z_2,\kappa)=\tilde g^Q (z_1)\tilde g^Q(z_2)
\frac{\rho_b\big(1-\rho_b\big)}{\left(1-\lambda_Q\alpha^Q(\kappa)\right)}
\label{DeltaF}
\end{equation}
where $\tilde g^Q(z)=\rho'^Q(z)/\left|\int dz'\;\rho'^Q(z')\right|$.
The first term of (\ref{DeltaF}) can be interpreted as a normalized function which constrains the range of the 
correlation function to the interface, i.e. where $\rho'_Q(z)$ does not vanish. The factor $\rho_b\big(1-\rho_b\big)$
guarantees the correct $\rho_b\to 0$ limit preserving the symmetry of the model. 
The chosen form of $\Delta F^Q$ (\ref{DeltaF}) reproduces the momentum dependence of the 
known Random Phase Approximation \cite{HansenMacDonald} in which $\alpha^Q(\kappa)$
 plays the role of an effective interaction potential on the interface. The yet unknown 
parameter 
$\lambda_Q$ is determined via the surface compressibility sum rule (\ref{surfcompsumrule}), 
valid for each $Q$-system.
By use of the parametrization (\ref{surfun}) in the sum rule (\ref{surfcompsumrule}) we obtain the
explicit relationship between $\lambda_Q$ and the thermodynamics:
\begin{equation}
\frac{\partial^2\left(-\beta\sigma^Q\right)}{\partial(\beta\mu)^2}={\left|\int dz\;\rho'_Q(z)\right|^2}
F_s(\kappa=0) = \frac{\rho_b(1-\rho_b)} {1-\lambda_Q\alpha^Q(\kappa=0)}
\label{h1} 
\end{equation}
Substituting the approximation of $F^Q_s(\kappa)$ in the evolution equation (\ref{latticeeq})
we obtain a closed non-linear partial differential equation for the surface free energy $\sigma^Q(\mu)$ at fixed
temperature. The initial condition for this equation is given by the mean field theory. 
The boundary conditions are given at coexistence {\em i.e.} $\mu=\mu_{co}$ where we assume that the surface susceptibility 
diverges (wetting) and at $\mu\to -\infty$ which corresponds to vanishing bulk density. The equation is written
in quasi-linear form by a suitable change of variable shown in Appendix B.
The evolution equation is then solved by a predictor-corrector algorithm \cite{Ames}. 

\begin{figure}[ht]
\includegraphics[scale=0.8,angle=270]{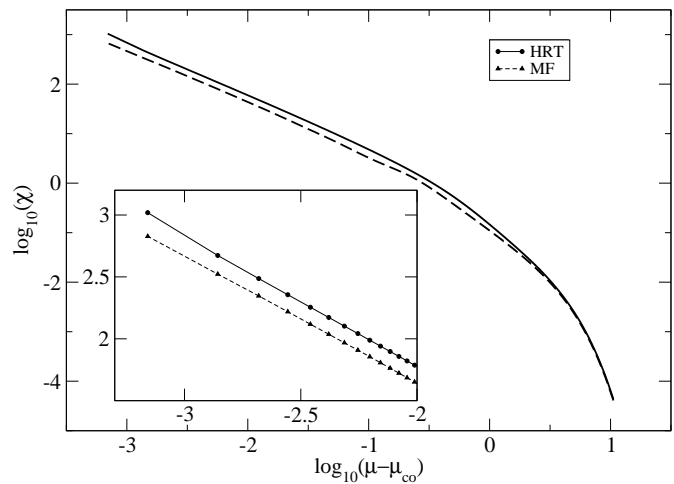}
\caption{\small The double logarithm plot of $\chi_s$  isotherm for $\beta=v/k_bT=0.72$ and $u/v=0.95$. We see that the effect of fluctuation is a renormalization of the critical adsorption amplitude.}
\label{chi72}
\end{figure}

\begin{figure}
\begin{center}
\includegraphics[scale=0.8,angle=270]{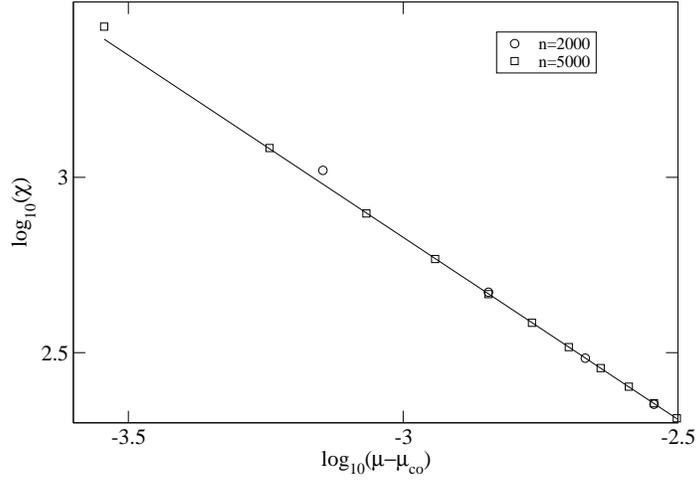}
\end{center}
\caption{\small The double Logarithmic plot of the $\chi_s$  isotherm for $\beta=v/k_bT=0.70$ and $u/v=0.95$ for two different numbers of mesh points. We see that the number of mesh points,{\em i.e.} the grid spacing, affects the value of $\chi_s$ at the last point of the grid. The line is only a guide to eye}
\label{chi70grid}
\end{figure}

\begin{figure}[ht]
\includegraphics[scale=0.8,angle=270]{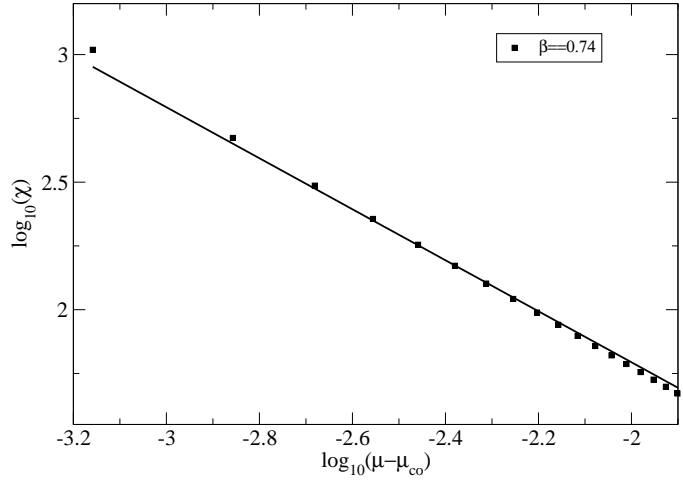}
\caption{\small The double logarithm plot of $\chi_s$  isotherm for $\beta=v/k_bT=0.72$ and $u/v=0.95$ in the asymptotic region. The line is  $\log(\chi_s)=-\log\big(\mu-\mu_{co}\big)+\log A$, with $A= 0.62$.}
\label{chi72fit}
\end{figure}

\begin{figure}[ht]
\includegraphics[scale=0.8,angle=270]{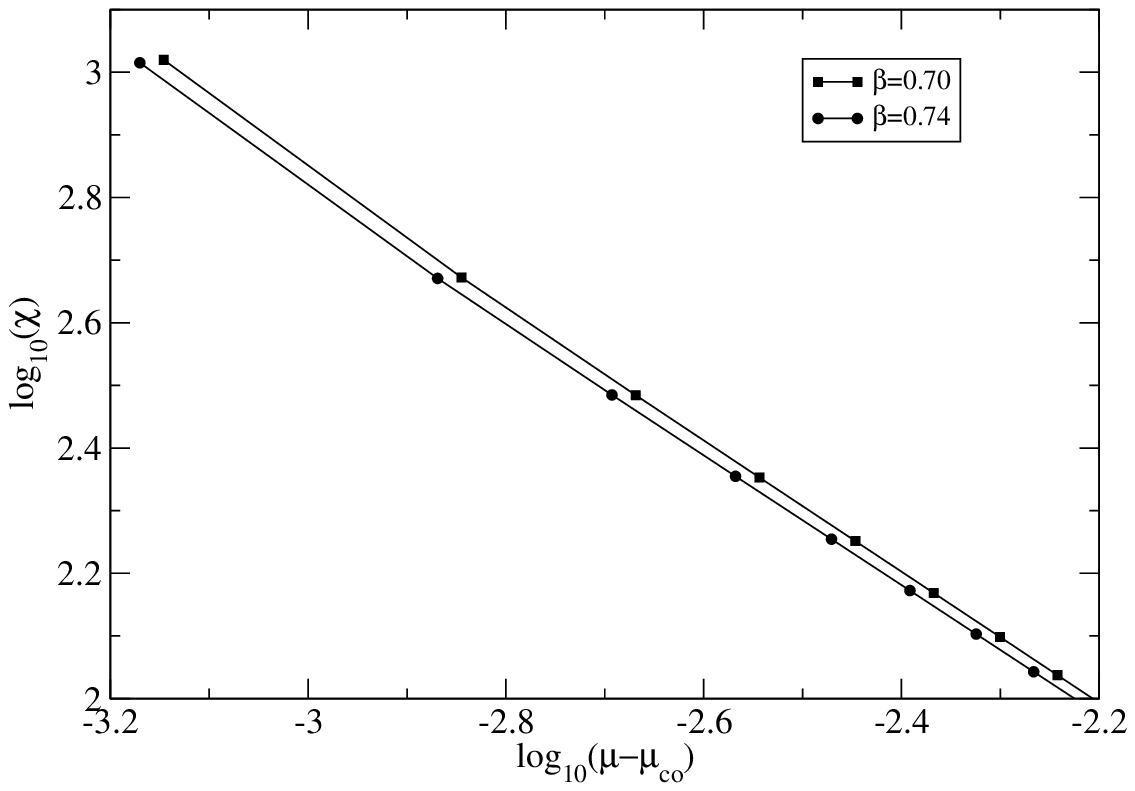}
\caption{\small The double logarithm plot of $\chi_s$  isotherm for $\beta=v/k_bT=0.70$ and $\beta=0.74$; $u/v=0.95$. 
We see that the effects of temperature on the critical amplitude is small. The lines are only a guides to eye}
\label{chi7074}
\end{figure}

In Fig.\ref{chi72} we show the logarithm of the susceptibility versus the logarithm of $\mu-\mu_{co}$. Notice that the value of
$\chi_s$ at the last point of the grid depends on the mesh spacing as can be seen in Fig.\ref{chi70grid}. For this reason this
value is not considered in the following analysis. In the asymptotic region, close to the coexistence curve, 
$\log(\chi_s)$ displays a linear behavior. We fit our data with a linear relation: $\log(\chi_s)\simeq a \log(\mu-\mu_{co})+b$ and 
we obtain $a=-1.040\pm 0.001$. This value is compatible with a logarithmic divergence of the adsorption {\em i.e.} the inclusion 
of fluctuations does not modify divergence of the adsorption so, as predicted by RG approach, the critical exponent for complete wetting
remain mean field like even when we introduce the fluctuations.  The main effect of fluctuations in the asymptotic region is 
rather a renormalization of the amplitude $A$, see Fig.\ref{chi72}. 
To estimate the value of the amplitude we fit the HRT results with $\log(\chi_s)=\log(A/\Delta\mu)=-\log\big(\mu-\mu_{co}\big)+\log A$, 
see Fig.\ref{chi72fit}. We obtain $A\simeq0.62$ which is larger than the lattice mean field result $A\simeq 0.45$. 
In Fig.\ref{chi7074} we plot the surface susceptibility for two different temperatures. 
The fluctuations introduce a weak dependence of the critical amplitude on the temperature, which is absent in mean-field results.
Using eq. (\ref{Aomega}) and our numerical result for $A$, we obtain $\omega_{HRT}\simeq 0.6$ 
which is smaller than the RG estimation for the Ising model $\omega\simeq 0.8$ \cite{EHP} and larger  
than the value $\omega\simeq 0.3$ obtained from simulation results on critical exponents for critical wetting \cite{BinderParry}.
However our numerical estimate is obtained at higher temperature: $T\simeq 0.92 T_c$. The deficiencies of the
mean filed approximation prevents us to study a wide range of temperature and so it is also difficult to extract 
from our results how $\omega$ depends on temperature. However our data seem to imply an increasing of the wetting parameter for 
$T\rightarrow T_c$ (see. Fig.\ref{chi7074}). 

\section{Conclusions}
In this work we have presented an extension of the HRT approach \cite{HRT} to inhomogeneous systems and we applied the
formalism to the wetting transition in the 3D Ising model in planar geometry. 
The approach presented here is, to our knowledge, the first {\sl microscopic} theory
able to describe capillary wave fluctuations at the level of the Renormalization Group 
approach. Its use is not restricted to the study of wetting but more generally to the study of
fluctuations effects in non-homogeneous systems. Differently from the standard bulk HRT we introduce a cylindrical cut-off 
to take into account the anisotropic spectrum of the fluctuations in such a reduced symmetry. We obtain a hierarchy of evolution 
equations for the bulk properties which reproduces the known RG approach close to the bulk critical point 
together with an additional hierarchy of differential equations for the surface free energy and the interfacial
correlations. This further hierarchy generally depends on bulk correlation functions. In order to close
the hierarchy at the lowest level, i.e. keeping only the free energy equations, we analyzed a simple 
Ornstein-Zernike ansatz for both the bulk and the interfacial correlations. Within this approximation, 
the resulting bulk critical exponents 
turn out to be exact to first order in $\epsilon=4-d$  and the surface free energy equation is 
equivalent to the known functional RG approach applied to the effective capillary wave Hamiltonian
\cite{fisher,lipowsky1,lipowsky}. 
To study the effects of fluctuations also in the non asymptotic region we investigated by the HRT approach 
the complete wetting transition in a lattice gas model with nearest-neighbor interaction (i.e. for the Ising model).  
In order to simplify the analysis we introduced a further decoupling approximation 
which disregards the effects of bulk fluctuations on surface quantities. The numerical results 
show an interesting renormalization of the critical adsorption amplitude which can be related to a renormalization 
of the wetting parameter. The HRT value of the wetting parameter is $\omega\simeq 0.6$  which is smaller the the 
field-theoretical estimate for this model \cite{EHP} but it is greater than simulation results for critical exponents of 
the critical wetting transition \cite{BinderParry} at lower temperature. This can probably related to the increase of
wetting parameter for $T\rightarrow T_c$.  
Notwithstanding the crudeness of some approximation we introduced,
the results for critical and non-critical quantities compare quite favorably with available simulation data.
Clearly there is room for improvement within the class of OZ closures of the HRT hierarchy. In particular
it would be interesting to allow for a renormalization of the transverse  correlation length $\xi_\perp$ which we kept 
constant. The extension of this approach to different systems, like off-lattice fluids or the Ising model in parallel plate 
geometry looks also promising. 
\acknowledgments
We thank  R. Evans for helpful suggestions.

\section*{APPENDIX A}
Consider a lattice system  with nearest-neighbor interaction
$$ v_{ij}=\left\{\begin{array}{lll} v & i,j\hbox{ nearest-neighbors site} \\
0 & \hbox{otherwise.} \end{array}\right.. $$
The Fourier transform of the dimensionless inter-particle potential is 
$$\phi(q,\kappa)=4 w \beta \big[\gamma_2(\kappa)+\cos(q)] $$
where $w=-v>0$ and 
$$\gamma_d(\kappa)=\frac{1}{d}\sum_{i=1}^d \cos(k_i).$$
The Brillouin zone is defined by $-\pi < k_i \le \pi$ and we choose the
cylindrical of the form \cite{HRT,HRTlattice}
$$\phi^Q(q,\kappa)=\left\{\begin{array}{ll} \phi(\kappa, q) 
& \gamma_2(\kappa)\leq  Q \\ 0 & \gamma_2(\kappa)>Q \end{array}\right.$$
with $Q\in [-1,1]$. The evolution equation for the surface tension of the lattice gas 
in three dimensions then becomes 
\begin{equation}
\frac{\partial \left(-\beta\sigma^Q\right)}{\partial Q}=\frac{1}{2}D_2(Q)\log\left[1+F_s(Q)\alpha(Q)\right] 
\end{equation}
where $\alpha(Q)$ is the value of $\alpha^Q(\kappa)$ evaluated 
at $\gamma_2(\kappa)=Q$ and $D_2(Q)$ is the two dimensional density of states:
$$ D_2(Q)=\int \frac{d^2\kappa}{(2\pi)^2}\;\delta(Q-\gamma_2(\kappa))=\frac{2}{\pi^2}K(\sqrt{1-Q^2}) $$
where $K(x)$ is complete elliptic integral of the first kind.

\section*{APPENDIX B} 
In this appendix we provide some detail on the algebraic manipulations on Eq. (\ref{latticeeq}) necessary for 
implementing an efficient numerical algorithm. 
As a first step we give the explicit form for the effective surface potential obtained by
substituting the parametrization discussed in the text for the function $g^Q(z)$ (\ref{gq}) 
into the definition of $\alpha^Q(\kappa)$ (\ref{alpha1}).
By use of the mean field density profile (\ref{mfr}) 
$g^Q(z)$ is just a Gaussian function and $\alpha^Q(\kappa)$ is independent of $l_Q$:
\begin{eqnarray}
&&\alpha^Q(\kappa)\simeq\frac{\big(\rho_b-\rho_1)^2}{2\pi}\int_{-\pi}^{\pi}dq\;
e^{-\frac{1}{2}q^2\xi_\perp^2}\phi(\kappa,q)=\frac{\big(\rho_b-\rho_1)^2}{2\pi}\int_{-\pi}^{\pi}dq\;
e^{-\frac{1}{2}q^2\xi_\perp^2}2\beta w \left[(d-1)\gamma(\kappa)+\cos(q)\right]= \nonumber \\
&&=\frac{\beta w (\rho_1-\rho_b)^2 \left[(d-1)I_1\gamma(\kappa)+I_2\right]}{\pi} 
\end{eqnarray}
where
\begin{eqnarray}
I_1&=&\int_{-\pi}^{\pi}dq\;e^{-\frac{1}{2}q^2\xi_\perp^2} \nonumber\\
I_2&=&\int_{-\pi}^{\pi}dq\;e^{-\frac{1}{2}q^2\xi_\perp^2}\cos(q) \nonumber
\end{eqnarray}
Now we introduce the shorthand notation 
$f_r=\rho_b\big( 1-\rho_b\big)$ and $b_Q=-\beta \sigma^Q$.  
The surface susceptibility of the $Q$-system is easily expressed in terms of $b''_Q$:
$$ b''_Q=\beta^2\frac{\partial^2 (-\beta\sigma_Q)}{\partial (\beta\mu)^2}=
\beta^2\frac{\partial\Gamma}{\partial \beta\mu}=\beta \chi_s $$
while the parameter $\lambda_Q$ is given by eq. (\ref{h1}).
The evolution equation (\ref{latticeeq}) is written in quasi-linear form by 
introducing the new variable  $v_Q$ defined by:
$$v_Q=\log\big(1+P(Q)\big)$$
where
$$P(Q)=F_s(Q)\alpha(Q)=\frac{b''_Q(\rho_b-\rho_1)^{-2}f_r\alpha_0\alpha(Q)}
{b''_Q\big(\alpha_0-\alpha(Q)\big)+f_r\beta^2\alpha(Q)}=\frac{b''_Qf_rr(Q)}{b''_Qm(Q)+f_r\beta^2\tilde{\alpha}(Q)} $$
and
$$ v_Q=f_r\tilde{\alpha}(Q)+\tilde{\alpha}^2(Q)u_Q. $$
where $\tilde{\alpha}(Q)=\alpha^Q(Q)(\rho_b-\rho_1)^{-2}$. By differentiating twice the
evolution equation (\ref{latticeeq}) with respect to the chemical potential and
substituting the definition of $v_Q$ we get
\begin{eqnarray}
\frac{\partial b''_Q}{\partial Q}=\frac{1}{2}D_2(Q)\frac{\partial^2 v_Q}{\partial \mu^2} \nonumber 
\end{eqnarray}
Inverting $v=v(b''_Q)$ we find
$$b''_Q=-\frac{\beta^2f_r\tilde{\alpha}(Q)\big[e^v-1\big]}{\big[\tilde{\alpha}_0-\tilde{\alpha}(Q)\big]
\big[e^v-1\big]-f_r\tilde{\alpha}_0\tilde{\alpha}(Q)} $$
Differentiating this  relation with respect to $Q$ we obtain 
\begin{eqnarray}
&&\frac{\partial b''_Q}{\partial Q}=-\frac{\partial }{\partial Q}
\left(\frac{\beta^2 f_r\tilde{\alpha}(Q)\big[e^{v_Q}-1\big]}{\big[\tilde{\alpha}_0-\tilde{\alpha}(Q)\big]
\big[e^{v_Q}-1\big]-f_r\tilde{\alpha}_0\tilde{\alpha}(Q)}\right)=\frac{\beta^2f^2_r\tilde{\alpha}(Q)^2
\tilde{\alpha_0}\dot{v}_Qe^{v_Q}-\beta^2f_r\dot{\tilde{\alpha}}\tilde{\alpha}_0\big[e^{v_Q}-1\big]^2}
{\left(m(q)\big[e^{v_Q}-1\big]-f_r r\right)^2} 
\label{monster}
\end{eqnarray}
where $\dot{v}_Q=\frac{\partial v_Q}{\partial Q}$, $v''=\frac{\partial^2 v_Q}{\partial \mu^2}$.
We now introduce the notation
\begin{eqnarray}
&&r(Q)=\tilde{\alpha}_0\tilde{\alpha}(Q)=\left(\frac{\beta}{\pi}\right)^2\left[(d-1)^2I_1^2Q+I_1I_2(d-1)(1+Q)+I_2^2\right]
\nonumber \\
&&m(Q)=\tilde{\alpha}_0-\tilde{\alpha}(Q)=\frac{\beta}{\pi}(d-1)I_1(1-Q) \nonumber
\end{eqnarray}
which give $\dot{r}=\tilde{\alpha}_0\dot{\tilde{\alpha}}$ and $\dot{m}=-\dot{\tilde{\alpha}}$. 
By solving the algebraic equation (\ref{monster}) for $\dot{v}_Q$ we obtain
\begin{eqnarray}
&&\dot{v}_Q=\frac{\dot{\tilde{\alpha}}}{f_r\tilde{\alpha}^2(Q)}
\big[e^{v_Q}-1\big]^2e^{-v_Q}+\frac{D_2(Q)e^{-v_Q}}{2\beta^2f^2_r\tilde{\alpha}^2(Q)\tilde{\alpha}_0}
\left(m(Q)\big[e^{v_Q}-1\big]-f_r r\right)^2v_Q'' 
\label{vequation}
\end{eqnarray}
which is the evolution equation for $v_Q$. Finally, by a further change of variable, we introduce $u_Q$ by 
\begin{equation}
f_r\tilde{\alpha}(Q)+\tilde{\alpha}(Q)^2u_Q=\log\left(1+F_s(Q)\alpha(Q)\right)
\label{uq}
\end{equation}
Switching from the chemical potential to the fugacity $z=e^{\beta(\mu-\mu_{co})}$ we express the derivatives
of $v_Q$ in terms of $u_Q$: 
\begin{eqnarray}
&&\dot{v}_Q=f_r\dot{\tilde{\alpha}}+2\tilde{\alpha}\dot{\tilde{\alpha}}u_Q+\tilde{\alpha}^2\dot{u}_Q \nonumber \\
&&v''_Q=\tilde{\alpha}f_r''+\tilde{\alpha}^2u_Q''=\tilde{\alpha}\beta^2\left[z\frac{\partial f_r}
{\partial z}+z^2\frac{\partial^2 f_r}{\partial z^2}\right]+
\tilde{\alpha}^2\beta^2\left[z\frac{\partial u_Q}{\partial z}+z^2\frac{\partial^2 u_Q}{\partial z^2}\right] \nonumber
\end{eqnarray}
Substituting these expressions into the evolution equation for $v_Q$ (\ref{vequation}) 
we obtain the quasi-linear equation satisfied by $u_Q$: 
\begin{equation}
\frac{\partial u_Q}{\partial Q}=L_1+L_2+M\frac{\partial^2 u}{\partial z^2} \label{uequation}
\end{equation}
with
\begin{eqnarray}
&&L_1=\frac{1}{\tilde{\alpha}^2}\left[\frac{\dot{\tilde{\alpha}}}{f_r\tilde{\alpha}^2}
\big(e^{v_Q}-1\big)^2e^{-v_Q}-f_r\dot{\tilde{\alpha}}-2\tilde{\alpha}\dot{\tilde{\alpha}}u_Q\right]
\label{L1}\\
&&L_2=\frac{D_2(Q)e^{-v_Q}}{2\beta^2\tilde{\alpha}^4(Q)f^2_r\tilde{\alpha}_0}
\left(m(Q)\big[e^{v_Q}-1\big]-f_rr\right)^2\left(\tilde{\alpha}\beta^2\left[z\frac{\partial f_r}{\partial z}+
z^2\frac{\partial^2 f_r}{\partial z^2}\right]+\tilde{\alpha}^2\beta^2z\frac{\partial u_Q}{\partial z}\right)
\label{L2}\\
&&M=\frac{D_2(Q)e^{-v_Q}}{2\beta^2\tilde{\alpha}^2(Q)f^2_r\tilde{\alpha}_0}
\left(m(Q)\big[e^{v_Q}-1\big]-f_rr\right)^2\beta^2z^2
\label{M}
\end{eqnarray}

\end{document}